 \documentstyle[prc,twocolumn,aps]{revtex}
\begin{document}
\preprint{KFA-IKP(TH)-1994-39}
\draft
\title{
Combined Description of $\bf{\overline{N}N}$ Scattering and Annihilation
With A Hadronic Model
}
\author{
V. Mull$^{1)}$ and K. Holinde$^{1),2)}$
}
\address{
$^{1)}$ Institut f\"ur Kernphysik(Theorie), Forschungszentrum J\"ulich GmbH,
 D-52425 J\"ulich, Germany \\
$^{2)}$Departamento de Fisica Teorica, Facultad de Ciencias Fisicas,
Universidad de Valencia, Burjasot (Valencia), Spain }
\date{\today}

\maketitle

\begin{abstract}
A model for the nucleon-antinucleon interaction is presented which is based on
meson-baryon dynamics. The elastic part is the $G$-parity transform of the Bonn
$NN$ potential. Annihilation into two mesons is described in terms of
microscopic baryon-exchange processes including all possible combinations of
$\pi,\eta,\rho,\omega,a_0,f_0,a_1,f_1,a_2,f_2,K,K^*$.  The remaining
annihilation part is taken into account by a phenomenological energy- and state
independent optical potential of Gaussian form. The model enables a
simultaneous
description of nucleon-antinucleon scattering and annihilation phenomena with
fair quality.
\end{abstract}


\section{Introduction}
Quantum Chromodynamics (QCD) is the theory of strong interactions with quarks
and gluons representing the fundamental degrees of freedom. Nevertheless, in
the
low energy regime, an effective theory in terms of collective, hadronic degrees
of freedom is probably the most efficient way to quantitatively describe most
strong interaction phenomena. In principle, the formulation and treatment of
QCD
can be done in terms of either the fundamental or the collective variables.  It
is a matter of convenience which set to choose under specific circumstances. Of
course, due to the enormous complexity of the theory, this issue is of decisive
importance when it comes to actual calculations.

Under this viewpoint, quark effects in low and medium energy physics have to be
defined as phenomena which cannot be understood in terms of only a few hadronic
variables but, on the other hand, have a simple quark-gluon interpretation. In
order to unambigously prove (or disprove) the existence of such signals in
nuclear physics it is essential to treat as many hadronic reactions as possible
from a conventional viewpoint, in terms of baryons and mesons. Only in this way
one will be able to reliably explore the limits of the conventional framework
and possibly establish discrepancies with the empirical situation, which might
then be identified with explicit quark-gluon effects.

Reactions involving antinucleons (for a review see e.g.\ the papers by Amsler
and Myhrer \cite{AM} and Dover et al.\cite{DGMF}) have always been considered
to
be the ideal place for finding quark effects since annihilation phenomena from
the nucleon-antinucleon (${\overline{N}N}$) system are supposedly governed by
short-distance physics.

There is general consensus however that, for large and medium distances
($r>1fm$) the elastic $\overline{N}N$ interaction is well described in terms of
meson exchanges and can be reliably obtained from a $G$-parity transform of
suitable $NN$ models. On the other hand, for short distances, there is at
present no reliable theory in this sector; therefore, the common attitude
(taken
e.g.\ by the Nijmegen\cite{Nijmegen} and Paris\cite{Paris} group) is to content
oneself with a phenomenological parameterization of this region. Both groups
have about 30 parameters at their disposal and obtain impressive fits to the
wealth of existing experimental data. The hope (expressed by the Paris group)
is
that the short range $\overline{N}N$ interaction so determined provides
``valuable hints in the elaboration of a deeper theoretical model''
\cite{Paris}.

One should realize of course that such a method can only provide constraints
but
no unique answer. First of all, there are still differences and ambiguities in
the medium range part of the $G$-parity transformed potentials: different $NN$
potentials can be used to start from; there are uncertainties due to missing
contributions (like e.g. correlated $\rho\pi$ exchange \cite{NNpirho}) and
vertex form factor effects, which all reach out up to $1.5fm$ or so. All these
topics should affect the result for the short range piece. Moreover, the usual
parameterization in Ref.\ \cite{Paris} assumes a very restricted non-locality
structure. Consequently we believe that a reliable test of a microscopic model
can only be made by confronting it to the experimental data directly.

The development of a dynamical model for the short range region is undoubtedly
a
formidable challenge, but it has to be met if we want to learn something about
the short-range dynamics and not give up from the beginning. Furthermore, in
order to prove the relevance of quark effects, it is not sufficient to
construct
a quark-gluon model which reasonably well describes the empirical situation. In
addition, the (possible) breakdown of the conventional hadronic picture, well
established in the outer range part, has to be shown by pointing out specific
discrepancies between model predictions and empirical data. This is precisely
the
motivation for our studies of the $\overline{N}N$ sector, in the conventional
framework, which we began almost 10 years ago.

Clearly, the goals of such a program are completely different from those
typically advocated by the Nijmegen and Paris groups\cite{Nijmegen,Paris}. For
us, the main aim is to test, without any bias, a conventional dynamical model
for the short range part. Thus it would be even counterproductive to introduce
sufficient parameters in order to obtain a quantitative fit to $\overline{N}N$
data since it would inhibit any conclusions about the physical relevance of the
model. Also, it is obviously essential to treat the short-range piece of both
the elastic and annihilation interaction in a consistent scheme.

Throughout, we use the $G$-parity transformed (full) Bonn $NN$ potential
\cite{MHE} as elastic $\overline{N}N$ interaction. This interaction has the
advantage (essential for our purpose) that it is prescribed everywhere, i.e.\
for (arbitrarily) short distances. Thus we are not forced to introduce any ad
hoc
parameters and in this way lose the predictiveness of our model from the
beginning, which would surely reduce (if not destroy completely) the
possibilities for a serious test of annihilation mechanisms.

Indeed, a good description of empirical $\overline{N}N$ (elastic as
well as charge-exchange) scattering data can be achieved with this
model by adding a simple phenomenological, state- and energy
independent optical potential with only 3 parameters to account for
annihilation (model A(BOX)\cite{NNbI}). Thus, no arbitrary adjustment
of the inner elastic part is a priory necessary in order to describe
the empirical data; obviously, this $G$-parity transform automatically
provides the spin, isospin and energy dependence phenomenologically
required. This is an important finding in itself.

Turning things around, this elastic $\overline{N}N$ interaction requires
essentially no state, isospin or energy dependence in the imaginary part and
thus seems to support an absorptive disk picture as dominant annihilation
mechanism. However, this cannot be the end of the story: In order to come to
reliable conclusions, it is essential to treat both the elastic and
annihilation
part of the interaction consistently in the same microscopic framework.

It is clear that due to the complexity of the $\overline{N}N$ annihilation
channels this program becomes quite involved and can only be pursued in steps.
In
Ref.\ \cite{NNbI} we have started by evaluating a selected set of two-meson
annihilation diagrams, $\overline{N}N\to M_1M_2\to\overline{N}N$, proceeding
via
baryon-exchange. All combinations of those mesons whose exchanges are
considered
in the elastic $\overline{N}N$ interaction (i.e.\
$\pi,\rho,\omega,\sigma,\delta$) have been included (with the same coupling
constants) as well as the strange mesons $K$ and $K^*$ generated by hyperon
exchange (with corresponding coupling constants taken from our hyperon-nucleon
model \cite{Holz}).

However, since these channels account at most only for about 30\% of the total
annihilation, their contributions have been artificially enhanced in Ref.\
\cite{NNbI} in order to provide the total empirical annihilation rate. (This
has
been achieved by suitably adjusting form factor parameters.)  This procedure of
using only a few annihilation channels leads to a very pronounced state and
isospin
dependence of the annihilation interaction because of strict selection rules
for
each annihilation process as a consequence of the conservation of isospin,
total
angular momentum, parity and $G$-parity. This model (called $C$ in
Ref.\cite{NNbI}) represents therefore the other extreme compared to the state
independent model $A(BOX)$.  As shown in Ref.\cite{NNbI}, it actually fails to
describe the empirical ${\overline{N}N}\to{\overline{N}N}$ data quantitatively.
Obviously, the state and isospin dependence of the corresponding annihilation
interaction is too strong. (Certainly we could have improved the fit
considerably by arbitrarily modifying the inner part of both the elastic and
annihilation interaction. Such a procedure, however, would completely obscure
the message and as we hope to have made clear, would be against the spirit of
our approach.)

In a second step \cite{NNbII} we have predicted $\overline{N}N\to M_1M_2$
transition rates going via baryon-exchange with adjustable form factor
parameters. For this, a DWBA procedure has been applied, with A(BOX)
and C as initial state interaction and no final (meson-meson)
interaction. A reasonable agreement with the experimental situation
could be achieved. However, there is a serious drawback of such a DWBA
approach: The annihilation which occurs both in the initial state
interaction and in the final transition process, is treated
inconsistently.

A consistent treatment can best be done in a coupled channels
framework, which yields both $\overline{N}N\to\overline{N}N$
and $\overline{N}N\to M_1M_2$ amplitudes at the same time. Liu and Tabakin
\cite{TabakinLiu} demonstrated the need for such an approach in $\overline{N}N$
physics and were the first to apply this method for explicit mesonic
channels, namely $\pi\pi$ and $\overline{K}K$. With about 20
parameters (used to parameterize further effective channels and the
short-range elastic part) they obtained a good simultaneous
description of (elastic and charge-exchange) $\overline{N}N$ scattering as well
as $\overline{N}N\to\pi\pi,\overline{K}K$ annihilation data.

In this paper, we present a consistent model describing
$\overline{N}N$ scattering and annihilation into specific channels at
the same time, along the same lines. Compared to Ref.\
\cite{TabakinLiu} the set of explicitly included meson channels is
considerably enlarged. Namely, apart from the pseudoscalar mesons
$\pi,\eta,K$ we consider all possible combinations of the lowest mass
mesons with $0^{++},1^{--},1^{++},2^{++}$ quantum numbers for both
isospin $I=0$ and $I=1$. This enlarged set of quantum numbers included
and the fact that all two-meson channels are now employed with a
realistic strength (in agreement with experimental information of
annihilation) turns out to strongly reduce the state dependence of the
annihilation interaction compared to our former model C and to
decisively improve the description of the data, as will be
demonstrated below.

 Apart from further two-meson channels with combinations of mesons
not considered so far, there is a part remaining which could be made up by
explicit three-meson channels or possible exotic contributions (glue-balls,
hybrids,\dots); In the model to be presented, this part is taken into account
by
a phenomenological optical potential of similar form as used in model $A(BOX)$
of Ref.\cite{NNbI}, but of course with modified parameters since part of the
annihilation is described microscopically.

The physical strength of the annihilation channels is determined by evaluating
all ${\overline{N}N}\to M_1 M_2$ cross sections (at rest and in flight) and
adjusting the form factor parameters at the annihilation vertices, which occur
in both the initial state interaction and the final ${\overline{N}N}\to M_1M_2$
transition, to available empirical information.  This consistency in the choice
of parameters is one major advantage compared to the former calculation
\cite{NNbII}, in which model $A(BOX)$ or $C$ has been used as initial state
interaction.

In the next section, we describe our model for ${\overline{N}N}$ scattering and
annihilation.  In section III we present and discuss the results and compare
these with our former models and experiment. Some concluding remarks are made
in
section IV.

\section{Model for $\bf {\overline{N}N}$ scattering and annihilation
into two mesons}

In principle, the microscopic treatment of the ${\overline{N}N}$ system is a
complicated problem involving couplings between various baryonic and mesonic
channels and diagonal interactions in all channels. In this paper we will
suppress any diagonal interaction except in the $\overline{N}N$ channel, the
reason being that not much is known about these interactions, especially in the
mesonic sector.  Also, this approximation is expected not severely to affect
the
main purpose of this work, which is to demonstrate that an increased number of
annihilation channels (with more meson quantum numbers $J^{PC}$) treated
explicitly
reduces the state dependence of the microscopic annihilation model and brings
in
this way the result towards the experiment. The coupled equations for the
$\overline{N}N$ scattering amplitude $T^{\overline{N}N\to\overline{N}N}$ and
the
transition amplitudes $T^{\overline{N}N\to M_1M_2}$ for the annihilation into
two
mesons can then be written as
\begin{eqnarray}
T^{\scriptscriptstyle {\overline{N}N}\to{\overline{N}N}} &{\enskip = \enskip} &
V^{\scriptscriptstyle{\overline{N}N}\to{\overline{N}N}} \enskip \nonumber \\
\label{ETNNb}
&& +\enskip V^{\scriptscriptstyle{\overline{N}N}\to{\overline{N}N}} \,
G^{\scriptscriptstyle{\overline{N}N}\to{\overline{N}N}} \,
T^{\scriptscriptstyle{\overline{N}N}\to{\overline{N}N}} \quad, \\
T^{\scriptscriptstyle {\overline{N}N}\to M_1 M_2}&{\enskip = \enskip}
&V^{\scriptscriptstyle {\overline{N}N}\to M_1 M_2} \nonumber \\
&&+V^{\scriptscriptstyle {\overline{N}N}\to M_1 M_2}\,
G^{\scriptscriptstyle{\overline{N}N}\to{\overline{N}N}} \,
T^{\scriptscriptstyle{\overline{N}N}\to{\overline{N}N}} \quad.
\label{ETNNbMM}
\end{eqnarray}
The ${\overline{N}N}$ interaction
$V^{\scriptscriptstyle{\overline{N}N}\to{\overline{N}N}}$ consists of an
elastic
and an annihilation part,
\begin{equation}
V^{\scriptscriptstyle{\overline{N}N}\to{\overline{N}N}}\enskip = \enskip V_{el}
\enskip + \enskip V_{ann} \quad .
\end{equation}

As stated in the Introduction, we use the $G$-parity transform of the (slightly
modified, cp. Ref. \cite{NNbI}) full Bonn $NN$ potential\cite{MHE} for that
purpose; corresponding diagrams are shown in Fig.\ \ref{FNNbInt}(a).
Compared to Ref.\ \cite{NNbI}, $V_{ann}$ is now split up into two parts
\begin{equation}
V_{ann} \enskip = \enskip \sum_{ij} V^{\scriptscriptstyle M_iM_j\to
        {\overline{N}N}}G^{\scriptscriptstyle M_iM_j} V^{\scriptscriptstyle
        {\overline{N}N}\to M_iM_j} \enskip + \enskip V_{opt} \ .
\label{EVANN}
\end{equation}

The first part results from a microscopic treatment of various two-meson
annihilation channels proceeding via baryon-exchange (Fig.\ \ref{FNNbInt}(b)).
It is important to realize that $V^{\scriptscriptstyle{\overline{N}N}\to
M_1M_2}$, which occurs both in the ${\overline{N}N}$ interaction (eq.\
(\ref{EVANN})) and in the actual annihilation process (eq.\ (\ref{ETNNbMM})) is
now treated in complete consistency.

The remaining contributions to the annihilation part of the ${\overline{N}N}$
interaction (involving e.g. the explicit transition into 3 and more mesons)
will
now be taken into account by an additional phenomenological piece
(fig. \ref{FNNbInt} (c)), for which we use the following parameterization in
coordinate space
\begin{equation}
V_{opt}\enskip =\enskip i\, W\, e^{-{r^2\over 2r_0^2}} \ .
\label{EVOPT}
\end{equation}
It is completely independent of energy, spin and isospin, with two parameters
($W= -1GeV, r_0 =0.4fm$) adjusted to the ${\overline{N}N}\to{\overline{N}N}$
cross section data.

In our model, the sum over $i,j$ in eq.\ (\ref{EVANN}) goes over all possible
combinations of $\pi,\eta,\rho,\omega,a_0,f_0,a_1,f_1,a_2,f_2$ (via $N,\Delta$
exchange) and $K,K^*$ (via $\Lambda,\Sigma,Y^*$ exchange), cp.\ fig.\
\ref{FTraPot}.  In order to obtain the transition interactions
$V^{\scriptscriptstyle {\overline{N}N}\to M_1M_2}$ we start from interaction
Lagrangians given in appendix A. As in the Bonn potential, the corresponding
diagrams have been evaluated within time-ordered perturbation theory.  Explicit
expressions and details can be found in \cite{NNbI,NNbII}.

As far as parameters are concerned, part of the coupling constants occurred
already in the Bonn potential \cite{MHE} and the Juelich hyperon-nucleon model
\cite{Holz} and could be taken over. Thus they are identical to those used in
the elastic interaction.  Remaining coupling constants are chosen in line with
empirical information\cite{Dumb}, only those without any information had to be
fitted to the ${\overline{N}N}$ cross sections.  Furthermore, the vertex
functions contain form factors, parameterized in a monopole type form
\begin{equation} F(\vec p\,^2_\delta) \enskip = \enskip
\Biggl( {{\Lambda_\delta^2 - M_\delta^2} \over {\Lambda_\delta^2 + {\vec
p_\delta}^{\ 2}}}\Biggr)^n \quad .\label{EFFD}\end{equation}
with $\vec p_\delta$ ($M_\delta$) being the 3-momentum (mass) of the baryon
exchanged in $V^{\scriptscriptstyle {\overline{N}N}\to M_1M_2}$.

Note that these form factors used in the annihilation diagrams have to be
distinguished from those used in the elastic meson-exchange process (although
the same particles are involved at the vertex) since now the exchanged baryon
is
the essential off-shell particle. Therefore the form factor is needed in a
quite
different kinematic region. The parameter $\Lambda_\delta$ should depend on the
type of particles involved at the vertex. However, in order to reduce the
number
of free parameters, we allow $\Lambda$ to depend only on the type of the
exchanged baryon but not on the produced meson. The values actually used are
given in Table \ref{TI}; they have been fixed in a self-consistency procedure
to
reproduce empirical annihilation data, see section III.

We mention finally that the conservation of parity $P$ and $G$-parity $G$
(resp.\ charge conjugation $C$) results in the following conditions (here the
primed magnitudes refer to the two-meson system, the unprimed ones to the
${\overline{N}N}$ system):

\begin{eqnarray}
 (-1)^{L+S+I}  &=&\cases{G_i G_j & if $M_i$,$M_j$  \cr
                                 & are  $G$--Eigenstates  \cr
                (-1)^{L^\prime+S^\prime+I} &otherwise \cr}\cr\ \cr
 (-1)^{L+1}   &=& P_i P_j\, (-1)^{L^\prime}.
\label{AR1}
\end{eqnarray}
The resulting selection rules are presented graphically in Table \ref{TII}.

\section{Results and Discussion}

The model specified in the last section (called model $D$ in the following)
provides definite predictions for both the ${\overline{N}N}$ scattering and
annihilation amplitude (eqs. (\ref{ETNNb},\ref{ETNNbMM})). From these, cross
sections and spin observables can be obtained in a straightforward
way. Throughout we will compare the results of model $D$ with corresponding
results of the preceding models $A(BOX)$ and $C$ (Ref.\cite{NNbI}).

\subsection{${\overline{N}N}$ scattering}

Fig.~\ref{FNNbCS} shows the total, integrated elastic and charge exchange as
well as the annihilation cross sections for ${\overline{p}p}$ scattering. Model
$D$ as well as A(BOX) of Ref.\cite{NNbI} (which contains absolutely no isospin
dependence for the annihilation part) agree with the empirical data. In
contrast, the result of the effective microscopic model $C$ of Ref.\cite{NNbI}
is by about a factor of 2 too large in the charge-exchange cross section. This
is because the isospin dependence of this annihilation model is too strong,
being generated by only a few annihilation channels.

Results for the differential cross sections for elastic
(${\overline{p}p}\to{\overline{p}p}$) scattering are shown in Fig.~\ref{figIX}.
The differential cross section is essentially flat for low energies, while
already for moderate energies a strong forward peak is seen, which clearly
demonstrates the importance of higher partial waves. For the highest energy
considered here, a minimum is observed in the cross section because diffractive
effects become relevant.  Throughout, model $A(BOX)$ provides a good
description, the agreement with the data is still reasonable for the consistent
annihilation model $D$ but fails completely for the effective model $C$.  A
similar situation is found for the elastic polarization (Fig.~\ref{figX}).
Model
$D$ accounts for the basic structures at low energies although there are
deficiencies at higher energies.  A rather reasonable agreement can be achieved
with model $A(BOX)$, while model $C$ predicts the wrong sign.

The results of Model $D$ and model $A(BOX)$ for the differential cross section
of the charge-exchange reaction ${\overline{p}p}\to{\overline{n}n}$
(Fig.~\ref{figXI}) almost coincide at higher energies.  For low energies,
however, differences occur at backward angles.  It has already been noted in
the
discussion of the integrated cross section, that the effective annihilation
model $C$ cannot provide a description of this process due to a too strong
isospin sensitivity of the annihilation.

The description of the charge exchange polarization data (Fig.~\ref{figXII}) is
still unsatisfactory, especially at backward angles and higher energies.

At the end of this section we show also the results for some selected spin
observables at two different energies, both for the elastic (Fig.\
\ref{figXIII}) and the charge-exchange channel (Fig.\ \ref{figXIV}).  For the
elastic spin-transfer observable $D_{nn}$ a few data points have been measured
\cite{KunneII}, for the charge-exchange $D_{nn}$ data have been obtained
recently
\cite{Bradamante}.

In summary, the increased number of explicit mesonic channels included
consistently, with a realistic strength, has obviously improved the
state dependence of our baryon exchange annihilation model
considerably. On the other hand, the comparison with the
phenomenological annihilation model A(BOX) shows that important
physics is still missing.  Again, we certainly had the possibility to
improve the agreement between the model $D$ results and experimental
data e.g.\ by adding spin-dependent (spin-orbit, tensor) parts to the
optical potential (Eq.\ (\ref{EVOPT})). For the reasons already
discussed in the Introduction we resisted against this temptation.

\subsection{Annihilation into two mesons}

In the following section we will look at the results for specific annihilation
channels.

Most data for the annihilation into two mesons are obtained from the
annihilation at rest in liquid or gaseous hydrogen\cite{AM,DGMF}. Table
\ref{TIII} contains all the relative cross sections for the 53 annihilation
channels included in our consistent annihilation model D. In order to
demonstrate the influence of the initial state interaction we also show the
results (with unchanged transition potential parameters) when $A(BOX)$ is used
instead of $D$ as initial state interaction.  The largest contributions are
given by the combinations of two vector mesons.  Vector-pseudoscalar and
vector-axialvector combinations also provide sizable fractions of the total
annihilation. These findings are independent of the particular initial state
interaction model used and are determined by the relevant vertex structures in
the baryon exchange diagrams. Other channels or combinations are of minor
importance; however, they tend to weaken the state dependence of the total
annihilation, a feature obviously favored by the empirical data. The channels
marked with an asterix can be reached for an annihilation at rest only when the
width of the mesons is taken into account, because the sum of their rest masses
is larger than twice the nucleon mass.

The fit of these branching ratios can certainly be improved by relaxing the
condition that the cutoff mass $\Lambda$ in the
baryon-exchange diagrams does not depend on the meson produced. For example,
for
nucleon exchange, the value $\Lambda_N = 1.5 GeV$ is mainly determined from the
experimentally well known annihilation channels involving $\pi,\rho$ and
$\omega$. For spin-2 mesons ($a_2,f_2$) the required dipole-form ($n=2$) then
leads to a relative suppression, which could be counterbalanced by a higher
value of $\Lambda$. As seen from Table \ref{TIII}, this would bring the
theoretical
results in better agreement with experiment.

A great success of the recent experiments done at LEAR is the measurement of
branching ratios together with the determination of the quantum numbers of the
initial ${\overline{N}N}$ state\cite{DGMF}. Table \ref{TIV} shows ratios of
branching ratios for either the same initial ${\overline{N}N}$ state into
different annihilation channels or into the same channel from different initial
${\overline{N}N}$ states.  These ratios express so-called dynamical selection
rules; a famous example is the first one (``$\pi\rho$ puzzle'').  Those
annihilation channels which are in principle allowed by the fundamental quantum
number conservation do not occur with equal probability or a statistical
distribution; the rate obviously depends sensitively on the channels and the
involved dynamics.  Ratios like those shown in Table \ref{TIV} are often
supposed to minimize the effects of the initial (and final) state
interactions. However, as clearly seen from the table, there is a strong
sensitivity to whether and even which kind of initial state interaction is
included. It does not drop out even if ratios from the same partial wave are
considered. (Note that the numbers of Table \ref{TIII} increase by an order of
magnitude if calculated in Born approximation!).  Thus a consistent description
for the transition model, initial state interaction (and probably also final
state interaction) is required before one can seriously address the question
about which transition mechanism is preferred. Conclusions based on Born
approximation appear to be premature.

For the most important annihilation channels cross section data for the
annihilation in flight exist, see Fig.~\ref{figXV}, which illustrate the energy
dependence of the annihilation mechanism. Obviously, model $D$ leads to a
satisfactory overall description.

For the $\pi^+\pi^-$ and the $\overline{K}K$ channel some more sensitive
observables have been measured, too: the differential cross section and the
analyzing power \cite{Hasan}.  It has been shown in Ref.\cite{pipiFSI} that the
description of both these observables requires more effort: here the
interactions between the outgoing mesons (which are to some extent known in
this
case)  seem to be essential for the
reproduction of the experimentally observed features of the data.  Work is in
progress to do a coupled channels calculation for these annihilation channels
including also $\pi\pi\to\pi\pi$, $\pi\pi\to\overline{K}K$, and
$\overline{K}K\to\overline{K}K$ interactions and is this way try
to describe also these high quality angle-dependent data.

\section{
Concluding remarks
}

One of the main topics of current research is to identify the relevant degrees
of freedom in low- and medium-energy strong interaction physics. The
short-range
part of the $\overline{N}N$ interaction represents a particular challenge in
this
respect, due to the considerable complexities introduced by the coupling of
various mesonic channels. In order to come to reliable conclusions, a
consistent
description of not only $\overline{N}N$ scattering but also
annihilation phenomena in
specific mesonic channels is required, with full inclusion of initial- and
final
state interaction effects. Such a program can be best done in a coupled
channels
framework. Since it requires an enormous effort it can only be done in steps by
increasing the number of explicit channels and/or including more and more
diagonal mesonic interactions. In this procedure it is advisable to keep the
number of free parameters small in order to avoid fitting the data
quantitatively while still missing a lot of important physics in the model.

In this paper, we have presented a conventional ha\-dronic model, in
terms of meson and baryon exchange, which enables a simultaneous
prediction of $\overline{N}N$ scattering and annihilation phenomena in
the two-meson sector involving the lowest mass $J^P=0^\pm,1^\pm,2^+$
mesons for both isospin $I=0$ and $I=1$. Given that we have only about 10
energy-independent free parameters (some open coupling constants, 5
cutoff masses in the
baryon-exchange diagrams and 2 optical potential parameters) the
results presented are, in our opinion, already quite encouraging
proving at least that also in the $\overline{N}N$ sector the
conventional hadronic concept is worth to be pursued further and is
surely a valid alternative to quark-gluon models. Still, remaining
discrepancies to empirical data (especially in the spin observables)
are a reflection of the fact that important physics is still
missing. Apart from further mesonic channels, diagonal mesonic
interactions as well as direct couplings between the various mesonic
channels have to be included, which represents a challenging task for
the future.
\appendix
\section{Interaction Lagrangians}
\label{hamop}
\noindent

The following Lagrangians are used in
this work for the coupling of spin ${\scriptstyle {1\over 2}}$ baryons and
mesons:

\begin{eqnarray}%
{\cal L}_{\scriptscriptstyle BBS}  &{\enskip = \enskip} &
 {\textstyle g_{\scriptscriptstyle BBS}} \quad
 \overline{\Psi}_{\!\alpha} \,  \Psi_{\beta} \,  \Phi^j
\\
%
{\cal L}_{\scriptscriptstyle BBP}  &{\enskip = \enskip} &
  {\textstyle g_{\scriptscriptstyle BBP}\over m_{p}} \quad
 \overline{\Psi}_{\!\alpha} \enskip \gamma^5 \gamma^\mu \enskip
 \Psi_{\beta}\enskip \partial_\mu \Phi^j
\\
%
{\cal L}_{\scriptscriptstyle BBV}  &{\enskip = \enskip} &
 {\textstyle g_{\scriptscriptstyle BBV}} \quad
 \overline{\Psi}_{\!\alpha} \enskip \gamma^\mu \enskip
 \Psi_{\beta} \,  \Phi^j_\mu \nonumber \\
&&+
 {\textstyle f_{\scriptscriptstyle BBV}\over {4 M_{\scriptscriptstyle N}}}\quad
 \overline{\Psi}_{\!\alpha} \enskip \sigma^{\mu\nu} \enskip
 \Psi_{\beta} \enskip
 (\partial_\mu \Phi^j_\nu - \partial_\nu \Phi^j_\mu )
\\
%
{\cal L}_{\scriptscriptstyle BBA} &{\enskip = \enskip} &
 {\textstyle g_{\scriptscriptstyle BBA}} \quad
 \overline{\Psi}_{\!\alpha} \enskip \gamma^5 \gamma^\mu  \enskip
 \Psi_{\beta} \, \Phi^j_\mu
\\
%
{\cal L}_{\scriptscriptstyle BBT} &{\enskip = \enskip} &
 {\textstyle g_{\scriptscriptstyle BBT} \over M_{\scriptscriptstyle N}} \enskip
\left \{
 i \overline{\Psi}_{\!\alpha} \enskip
 \left (\gamma^\mu \partial^\nu \Psi_{\beta} +
        \gamma^\nu \partial^\mu \Psi_{\beta}  \right )
 \right . \nonumber \\ &&
\left .
 -i \left (\partial^\nu \overline{\Psi}_{\!\alpha}  \gamma^\mu +
           \partial^\mu \overline{\Psi}_{\!\alpha}  \gamma^\nu \right )
           \Psi_\beta
\right \}
   \enskip \Phi^j_{\mu\nu}
\end{eqnarray}

Vertices with a spin${\scriptstyle {1\over 2}}$-,
a spin${\scriptstyle {3\over 2}}$-Baryon and a meson are given by the
Lagrangians:

\begin{eqnarray}
{\cal L}_{\scriptscriptstyle BDP} &{\enskip = \enskip} &
  {\textstyle g_{\scriptscriptstyle BDP}\over m_{p}}
 \quad
 \overline{\Psi}_{\!\alpha} \, \Psi^\mu_{\beta}  \enskip
 \partial_\mu \Phi^j
\\
{\cal L}_{\scriptscriptstyle BDV} &{\enskip = \enskip} & {\textstyle
 g_{\scriptscriptstyle BDV} \over m_v} \quad \overline{\Psi}_{\!\alpha} \enskip
 i \gamma^5 \gamma^\mu \enskip \Psi^\nu_{\beta} \enskip \left (\partial_\mu
 \Phi^j_\nu - \partial_\nu \Phi^j_\mu \right )
\end{eqnarray}
with the following meaning of the indices:

\centerline{\vtop{
\halign{ $#$ &\ #\ & \  $#$&\ # \hfil\ &\ $#$  \cr
B:& spin ${\scriptstyle {1\over 2}}$-baryons   & \cr
D:& spin ${\scriptstyle {3\over 2}}$-baryons & \cr
S:& scalar   mesons      &J^P=0^+ \cr
P:& pseudoscalar mesons &J^P=0^- \cr
V:& vector mesons        &J^P=1^- \cr
A:& axialvector mesons   &J^P=1^+ \cr
T:& tensor mesons        &J^P=2^+ \cr
}}}

\pagebreak

\begin{figure}
\caption{
\label{FNNbInt}
{Elastic (a), microscopic annihilation (b), and phenomenological annihilation
(c)
part of our ${\overline{N}N}$ interaction model.}
}
\end{figure}

\begin{figure}
\caption{
\label{FTraPot}
Transition potentials $V^{\scriptscriptstyle{\overline{N}N}\to M_i M_j}$
included explicitly in our microscopic annihilation model.}
\end{figure}

\begin{figure}
\caption{
\label{FNNbCS}
{
Total, elastic, charge-exchange and annihilation
cross sections for ${\overline{p}p}$ scattering.
Results of the consistent model $D$ are given by the
solid lines, dashed lines correspond to the phenomenological
model $A(BOX)$ and dashed-dotted lines result from the
 effective microscopic model $C$.
The references for the data can be found in Ref.\protect\cite{DGMF,MullDiss}.
}}
\end{figure}

\begin{figure}
\caption{
\label{figIX}
Elastic ${\overline{p}p}$ differential cross sections at various energies.  The
data are taken from
Ref.\protect\cite{BruecknerIV,KiI,KunneI,Bertini,KageyamaI}.
The same description of the curves as in figure \protect\ref{FNNbCS}.  }
\end{figure}

\begin{figure}
\caption{
\label{figX}
Elastic ${\overline{p}p}$ polarizations at some energies.  The data are taken
from Ref.\protect\cite{BruecknerIV,KiI,KunneI,Bertini,KageyamaI}.  The same
description of the curves as in figure \protect\ref{FNNbCS}.  }
\end{figure}

\begin{figure}
\caption{
\label{figXI}
Charge-exchange differential cross sections.
The data are taken from Ref.\protect\cite{NakII,BruecknerII,Birsa}.
The same description of the curves as in figure \protect\ref{FNNbCS}.
}
\end{figure}

\begin{figure}
\caption{
\label{figXII}
Charge-exchange polarizations.
The data are taken from Ref.\protect\cite{NakII,BruecknerII,Birsa}.
The same description of the curves as in figure \protect\ref{FNNbCS}.
}
\end{figure}
\begin{figure}
\caption{
\label{figXIII}
Some spin observables in the elastic channel: $C_{nn},D_{nn}$ and $K_{nn}$ at
two different energies.  The data are taken from
Ref.\protect\cite{KunneII}.  The same description of the curves
as in figure \protect\ref{FNNbCS}.  }
\end{figure}

\begin{figure}
\caption{
\label{figXIV}
The spin observables $C_{nn},D_{nn}$ and $K_{nn}$ for the
charge-exchange reaction.
The data are taken from Ref.\protect\cite{Bradamante}.
The same description of the curves as in figure \protect\ref{FNNbCS}.
}
\end{figure}

\begin{figure}
\caption{
\label{figXV}
Annihilation cross sections ``in flight'' for the
most important channels.
The same description of the curves as in figure \protect\ref{FNNbCS}.
The data are calculated from values given in Ref.\protect\cite{BardinII,Sai}.
}
\end{figure}
\pagebreak

\begin{table}
\caption{\label{TI}
Coupling constants and cutoff parameters
in the transition potential $V^{{\overline{N}N}\to M_iM_j}$.}

{
\hbox{\hfill
\vtop{
\begin{tabular}{cddcc}

 Vertex & ${g^2\over 4\pi}$ &f/g& $\Lambda$& $n$   \cr
\noalign{\hrule }
$NN\pi$          & 0.0778&        &  1500 & 1  \cr
$NN\eta$         & 0.6535&        &  1500 & 1  \cr
$NN\rho$         & 0.84  & 6.1    &  1500 & 1  \cr
$NN\omega$       & 20.0  & 0      &  1500 & 1  \cr
$NN f_0$         & 5.723 &        &  1500 & 1  \cr
$NN a_0$         & 2.6653&        &  1500 & 1  \cr
$NN f_1$         & 10.0  &        &  1500 & 1  \cr
$NN a_1$         &  7.0  &        &  1500 & 1  \cr
$NN f_2$         & 2.0   &        &  1500 & 2  \cr
$NN a_2$         & 4.0   &        &  1500 & 2  \cr
$N\Delta\pi  $  & 0.224     &      &  1700 & 1  \cr
$N\Delta\rho $  & 20.45     &      &  1700 & 2  \cr
$N\Lambda K  $  &0.9063 &      &  1800 & 1  \cr
$N\Lambda K^*$  &2.5217 &-5.175&  1800 & 1  \cr
$N\Sigma  K  $  &0.0313 &      &  2000 & 1  \cr
$N\Sigma  K^*$  &0.8409 &2.219 &  2000 & 1  \cr
$N Y^*    K  $  &0.0372 &      &  2000 & 1  \cr
$N Y^*    K^*$  &3.4077 &      &  2000 & 2  \cr
\end{tabular}}
\hfill}}

\end  {table}

\begin{table}
\squeezetable
\caption{\label{TII}Conservation of parity and $G$-parity
impose selection rules on the transition from the ${\overline{N}N}$ system to
the
two meson system: all hashed fields are generally forbidden by parity
conservation. Transitions marked by the letter A are allowed for
meson pair $G$-parity
$G'=(-1)^{(I+J)}$, whereas transitions marked by B
can occur for $G'=(-1)^{(I+J+1)}$.
}
\medskip
\small
\ifpreprintsty
\tiny
\fi
\offinterlineskip
\dimen1=1.6cm
\dimen0=0.1cm
\dimen3=0.1cm

\halign{\strut \vrule# &\ \hfil # \hfil\ &\vrule# & \ $L'=#$\hfil\ & \ \hfil
$S'=#$\hfil\ & \vrule #&&\hbox to\dimen1{ \hfil # \hfil}&\vrule #\cr
\noalign{\hrule} &&&\multispan 2 \hbox{meson-state} &&\multispan 5
${\overline{N}N}$-state &\cr \noalign{\hrule} &&&\omit&\omit &&\ $L=J$\ &&\
$L=J$\ &&\ $L=J\pm1$\ & \cr &&&\omit&\omit &&\ $S=0$\ &&\ $S=1$\ &&\ $S=1$\ &
\cr &&&\omit&\omit &&\ Sing \ &&\ Trip \ &&\ Coup \ & \cr
\noalign{\hrule} &PP&& J & 0 &&\omit\hbox to \dimen1{\hskip 0.025cm
\leaders\hbox{\vrule width 0.025cm height8.5pt depth
4pt\hskip0.10cm}\hfill}&&\omit\hbox to \dimen1{\hskip 0.025cm
\leaders\hbox{\vrule width 0.025cm height8.5pt depth 4pt\hskip0.10cm}\hfill}&&
A
& \cr
\noalign{\hrule} &PS&& J & 0 && A && B && \omit\hbox to \dimen1{\hskip 0.025cm
\leaders\hbox{\vrule width 0.025cm height8.5pt depth 4pt\hskip0.10cm}\hfill}&
\cr
\noalign{\hrule} &SS&& J & 0 &&\omit\hbox to \dimen1{\hskip 0.025cm
\leaders\hbox{\vrule width 0.025cm height8.5pt depth
4pt\hskip0.10cm}\hfill}&&\omit\hbox to \dimen1{\hskip 0.025cm
\leaders\hbox{\vrule width 0.025cm height8.5pt depth 4pt\hskip0.10cm}\hfill}&&
A
& \cr
\noalign{\hrule} &VP&& J & 1 &&\omit\hbox to \dimen1{\hskip 0.025cm
\leaders\hbox{\vrule width 0.025cm height8.5pt depth
4pt\hskip0.10cm}\hfill}&&\omit\hbox to \dimen1{\hskip 0.025cm
\leaders\hbox{\vrule width 0.025cm height8.5pt depth 4pt\hskip0.10cm}\hfill}&&
A
& \cr \omit\vrule &&&\multispan{2}\dotfill&&\multispan{5}\hrulefill&\cr & &&
J\pm1 & 1 && A && B && \omit\hbox to \dimen1{\hskip 0.025cm
\leaders\hbox{\vrule
width 0.025cm height8.5pt depth 4pt\hskip0.10cm}\hfill}& \cr
\noalign{\hrule}

&VS&& J & 1 && A && B && \omit\hbox to \dimen1{\hskip 0.025cm
\leaders\hbox{\vrule width 0.025cm height8.5pt depth 4pt\hskip0.10cm}\hfill}&
\cr \omit\vrule &&&\multispan{2}\dotfill&&\multispan{5}\hrulefill&\cr & &&
J\pm1
& 1 &&\omit\hbox to \dimen1{\hskip 0.025cm \leaders\hbox{\vrule width 0.025cm
height8.5pt depth 4pt\hskip0.10cm}\hfill}&&\omit\hbox to \dimen1{\hskip 0.025cm
\leaders\hbox{\vrule width 0.025cm height8.5pt depth 4pt\hskip0.10cm}\hfill}&&
A
& \cr
\noalign{\hrule} &AP&& J & 1 && A && B && \omit\hbox to \dimen1{\hskip 0.025cm
\leaders\hbox{\vrule width 0.025cm height8.5pt depth 4pt\hskip0.10cm}\hfill}&
\cr \omit\vrule &&&\multispan{2}\dotfill&&\multispan{5}\hrulefill&\cr & &&
J\pm1
& 1 &&\omit\hbox to \dimen1{\hskip 0.025cm \leaders\hbox{\vrule width 0.025cm
height8.5pt depth 4pt\hskip0.10cm}\hfill}&&\omit\hbox to \dimen1{\hskip 0.025cm
\leaders\hbox{\vrule width 0.025cm height8.5pt depth 4pt\hskip0.10cm}\hfill}&&
A
& \cr
\noalign{\hrule} &VV&& J & 0 &&\omit\hbox to \dimen1{\hskip 0.025cm
\leaders\hbox{\vrule width 0.025cm height8.5pt depth
4pt\hskip0.10cm}\hfill}&&\omit\hbox to \dimen1{\hskip 0.025cm
\leaders\hbox{\vrule width 0.025cm height8.5pt depth 4pt\hskip0.10cm}\hfill}&&
A
& \cr \noalign{\vskip-1pt} \omit\vrule
&&&\multispan{2}\dotfill&&\multispan{5}\hrulefill&\cr & && J & 1 &&\omit\hbox
to
\dimen1{\hskip 0.025cm \leaders\hbox{\vrule width 0.025cm height8.5pt depth
4pt\hskip0.10cm}\hfill}&&\omit\hbox to \dimen1{\hskip 0.025cm
\leaders\hbox{\vrule width 0.025cm height8.5pt depth 4pt\hskip0.10cm}\hfill}&&
A
& \cr \noalign{\vskip-1pt} \omit\vrule
&&&\multispan{2}\dotfill&&\multispan{5}\hrulefill&\cr & && J\pm1 & 1 && A && B
&& \omit\hbox to \dimen1{\hskip 0.025cm \leaders\hbox{\vrule width 0.025cm
height8.5pt depth 4pt\hskip0.10cm}\hfill}& \cr \noalign{\vskip-1pt} \omit\vrule
&&&\multispan{2}\dotfill&&\multispan{5}\hrulefill&\cr & && J & 2 &&\omit\hbox
to
\dimen1{\hskip 0.025cm \leaders\hbox{\vrule width 0.025cm height8.5pt depth
4pt\hskip0.10cm}\hfill}&&\omit\hbox to \dimen1{\hskip 0.025cm
\leaders\hbox{\vrule width 0.025cm height8.5pt depth 4pt\hskip0.10cm}\hfill}&&
A
& \cr \noalign{\vskip-1pt} \omit\vrule
&&&\multispan{2}\dotfill&&\multispan{5}\hrulefill&\cr & && J\pm1 & 2 && A && B
&& \omit\hbox to \dimen1{\hskip 0.025cm \leaders\hbox{\vrule width 0.025cm
height8.5pt depth 4pt\hskip0.10cm}\hfill}& \cr \noalign{\vskip-1pt} \omit\vrule
&&&\multispan{2}\dotfill&&\multispan{5}\hrulefill&\cr & && J\pm2 & 2
&&\omit\hbox to \dimen1{\hskip 0.025cm \leaders\hbox{\vrule width 0.025cm
height8.5pt depth 4pt\hskip0.10cm}\hfill}&&\omit\hbox to \dimen1{\hskip 0.025cm
\leaders\hbox{\vrule width 0.025cm height8.5pt depth 4pt\hskip0.10cm}\hfill}&&
A
& \cr
\noalign{\hrule} &AV&& J & 0 && A && B && \omit\hbox to \dimen1{\hskip 0.025cm
\leaders\hbox{\vrule width 0.025cm height8.5pt depth 4pt\hskip0.10cm}\hfill}&
\cr \noalign{\vskip-1pt} \omit\vrule
&&&\multispan{6}\dotfill&&\omit\hrulefill&\cr & && J & 1 && A && B &&
\omit\hbox
to \dimen1{\hskip 0.025cm \leaders\hbox{\vrule width 0.025cm height8.5pt depth
4pt\hskip0.10cm}\hfill}& \cr \noalign{\vskip-1pt} \omit\vrule
&&&\multispan{2}\dotfill&&\multispan{5}\hrulefill&\cr & && J\pm1 & 1
&&\omit\hbox to \dimen1{\hskip 0.025cm \leaders\hbox{\vrule width 0.025cm
height8.5pt depth 4pt\hskip0.10cm}\hfill}&&\omit\hbox to \dimen1{\hskip 0.025cm
\leaders\hbox{\vrule width 0.025cm height8.5pt depth 4pt\hskip0.10cm}\hfill}&&
A
& \cr \noalign{\vskip-1pt} \omit\vrule
&&&\multispan{2}\dotfill&&\multispan{5}\hrulefill&\cr & && J & 2 && A && B &&
\omit\hbox to \dimen1{\hskip 0.025cm \leaders\hbox{\vrule width 0.025cm
height8.5pt depth 4pt\hskip0.10cm}\hfill}& \cr \noalign{\vskip-1pt} \omit\vrule
&&&\multispan{2}\dotfill&&\multispan{5}\hrulefill&\cr & && J\pm1 & 2
&&\omit\hbox to \dimen1{\hskip 0.025cm \leaders\hbox{\vrule width 0.025cm
height8.5pt depth 4pt\hskip0.10cm}\hfill}&&\omit\hbox to \dimen1{\hskip 0.025cm
\leaders\hbox{\vrule width 0.025cm height8.5pt depth 4pt\hskip0.10cm}\hfill}&&
A
& \cr \noalign{\vskip-1pt} \omit\vrule
&&&\multispan{2}\dotfill&&\multispan{5}\hrulefill&\cr & && J\pm2 & 2 && A && B
&& \omit\hbox to \dimen1{\hskip 0.025cm \leaders\hbox{\vrule width 0.025cm
height8.5pt depth 4pt\hskip0.10cm}\hfill}& \cr
\noalign{\hrule} &TP&& J & 2 && A && B && \omit\hbox to \dimen1{\hskip 0.025cm
\leaders\hbox{\vrule width 0.025cm height8.5pt depth 4pt\hskip0.10cm}\hfill}&
\cr \noalign{\vskip-1pt} \omit\vrule
&&&\multispan{2}\dotfill&&\multispan{5}\hrulefill&\cr & && J\pm1 & 2
&&\omit\hbox to \dimen1{\hskip 0.025cm \leaders\hbox{\vrule width 0.025cm
height8.5pt depth 4pt\hskip0.10cm}\hfill}&&\omit\hbox to \dimen1{\hskip 0.025cm
\leaders\hbox{\vrule width 0.025cm height8.5pt depth 4pt\hskip0.10cm}\hfill}&&
A
& \cr \noalign{\vskip-1pt} \omit\vrule
&&&\multispan{2}\dotfill&&\multispan{5}\hrulefill&\cr & && J\pm2 & 2 && A && B
&& \omit\hbox to \dimen1{\hskip 0.025cm \leaders\hbox{\vrule width 0.025cm
height8.5pt depth 4pt\hskip0.10cm}\hfill}& \cr
\noalign{\hrule} &TV&& J & 1 && A && B && \omit\hbox to \dimen1{\hskip 0.025cm
\leaders\hbox{\vrule width 0.025cm height8.5pt depth 4pt\hskip0.10cm}\hfill}&
\cr \noalign{\vskip-1pt} \omit\vrule
&&&\multispan{2}\dotfill&&\multispan{5}\hrulefill&\cr & && J\pm1 & 1
&&\omit\hbox to \dimen1{\hskip 0.025cm \leaders\hbox{\vrule width 0.025cm
height8.5pt depth 4pt\hskip0.10cm}\hfill}&&\omit\hbox to \dimen1{\hskip 0.025cm
\leaders\hbox{\vrule width 0.025cm height8.5pt depth 4pt\hskip0.10cm}\hfill}&&
A
& \cr \noalign{\vskip-1pt} \omit\vrule
&&&\multispan{2}\dotfill&&\multispan{5}\hrulefill&\cr & && J & 2 && A && B &&
\omit\hbox to \dimen1{\hskip 0.025cm \leaders\hbox{\vrule width 0.025cm
height8.5pt depth 4pt\hskip0.10cm}\hfill}& \cr \noalign{\vskip-1pt} \omit\vrule
&&&\multispan{2}\dotfill&&\multispan{5}\hrulefill&\cr & && J\pm1 & 2
&&\omit\hbox to \dimen1{\hskip 0.025cm \leaders\hbox{\vrule width 0.025cm
height8.5pt depth 4pt\hskip0.10cm}\hfill}&&\omit\hbox to \dimen1{\hskip 0.025cm
\leaders\hbox{\vrule width 0.025cm height8.5pt depth 4pt\hskip0.10cm}\hfill}&&
A
& \cr \noalign{\vskip-1pt} \omit\vrule
&&&\multispan{2}\dotfill&&\multispan{5}\hrulefill&\cr & && J\pm2 & 2 && A && B
&& \omit\hbox to \dimen1{\hskip 0.025cm \leaders\hbox{\vrule width 0.025cm
height8.5pt depth 4pt\hskip0.10cm}\hfill}& \cr \noalign{\vskip-1pt} \omit\vrule
&&&\multispan{6}\dotfill&&\omit\hrulefill&\cr & && J & 3 && A && B &&
\omit\hbox
to \dimen1{\hskip 0.025cm \leaders\hbox{\vrule width 0.025cm height8.5pt depth
4pt\hskip0.10cm}\hfill}& \cr \noalign{\vskip-1pt} \omit\vrule
&&&\multispan{2}\dotfill&&\multispan{5}\hrulefill&\cr & && J\pm1 & 3
&&\omit\hbox to \dimen1{\hskip 0.025cm \leaders\hbox{\vrule width 0.025cm
height8.5pt depth 4pt\hskip0.10cm}\hfill}&&\omit\hbox to \dimen1{\hskip 0.025cm
\leaders\hbox{\vrule width 0.025cm height8.5pt depth 4pt\hskip0.10cm}\hfill}&&
A
& \cr \noalign{\vskip-1pt} \omit\vrule
&&&\multispan{2}\dotfill&&\multispan{5}\hrulefill&\cr & && J\pm2 & 3 && A && B
&& \omit\hbox to \dimen1{\hskip 0.025cm \leaders\hbox{\vrule width 0.025cm
height8.5pt depth 4pt\hskip0.10cm}\hfill}& \cr \noalign{\vskip-1pt} \omit\vrule
&&&\multispan{2}\dotfill&&\multispan{5}\hrulefill&\cr & && J\pm3 & 3
&&\omit\hbox to \dimen1{\hskip 0.025cm \leaders\hbox{\vrule width 0.025cm
height8.5pt depth 4pt\hskip0.10cm}\hfill}&&\omit\hbox to \dimen1{\hskip 0.025cm
\leaders\hbox{\vrule width 0.025cm height8.5pt depth 4pt\hskip0.10cm}\hfill}&&
A
& \cr \noalign{\hrule} \noalign{\hrule} }
\end{table}

\begin{table}
\squeezetable
\caption{
\label{TIII}
Branching ratios ``at rest'' for 53 annihilation channels.
The data are taken from Ref.\protect\cite{AM,DGMF}.
}
\begin{tabular}{c|d |d |c}
$\overline{p}p  \to $         & D      & A(BOX)& EXP.\  \cr
\noalign{\hrule}
$\pi^+ \pi^-       $     & 0.54   & 0.92  &$0.33\pm0.017 $ \cr
$\pi^{\scriptscriptstyle 0} \pi^{\scriptscriptstyle 0} $     & 0.098
& 0.33  &$0.02-0.06    $ \cr
$\pi^{\scriptscriptstyle 0} \eta     $     & 0.0095 & 0.01  &$ 0.03\pm0.02 $
\cr
$\eta \eta         $     & 0.0037 &0.0075 &$0.008\pm0.003$ \cr
$\pi^\pm  a_0^\mp  $     & 0.013  &0.017  &$ 0.69\pm0.12$  \cr
$\pi^{\scriptscriptstyle 0} a_0^{\scriptscriptstyle 0} $     & 0.0021
&0.0046 &              \cr
$\eta     a_0^{\scriptscriptstyle 0} $     & 0.021  &0.014  &              \cr
$\pi^{\scriptscriptstyle 0} f_0      $     & 0.067  &0.035  &              \cr
$\eta f_0          $     & 0.0090 &0.024  &              \cr
$ a_0^+     a_0^-      $ & 0.0017 &0.0011 &              \cr
$ a_0^{\scriptscriptstyle 0}  a_0^{\scriptscriptstyle 0}   $ & 0.0007
&0.0005 &              \cr
$ f_0       a_0^{\scriptscriptstyle 0}   $ & 0.0002 &0.0002 &              \cr
$ f_0       f_0        $ & 0.0008 &0.0006 &              \cr
$\rho^{\pm}\pi^{\mp}$    & 2.32   & 3.94  &$ 3.4  \pm0.2 $ \cr
$\rho^{\scriptscriptstyle 0} \pi^{\scriptscriptstyle 0} $    & 0.85
& 1.21  &$ 1.4  \pm0.1 $ \cr
$\rho^{\scriptscriptstyle 0} \eta    $     & 0.09   & 0.64  &$ 0.65 \pm0.14$
\cr
$\omega\pi^{\scriptscriptstyle 0}    $     & 0.57   & 2.11  &$ 0.52 \pm0.05$
\cr
$\omega\eta        $     & 0.20   & 0.09  &$ 0.46 \pm0.14$ \cr
$\rho^\pm  a_0^\mp  $    & 0.63   & 0.40  &              \cr
$\rho^{\scriptscriptstyle 0} a_0^{\scriptscriptstyle 0} $    & 0.24
& 0.13  &              \cr
$\rho^{\scriptscriptstyle 0} f_0    $      & 0.90   & 1.82  &              \cr
$\omega a_0^{\scriptscriptstyle 0}  $      & 0.17   & 0.70  &              \cr
$\omega f_0       $      & 1.28   & 1.90  &              \cr
$a_1^{\pm}\pi^{\mp}$     & 1.03   & 0.83  &              \cr
$a_1^{\scriptscriptstyle 0} \pi^{\scriptscriptstyle 0} $     & 0.22
&0.17   &              \cr
$a_1^{\scriptscriptstyle 0} \eta   $       & 0.0078 &0.0088 &              \cr
$f_1 \pi^{\scriptscriptstyle 0}    $       & 0.47   & 0.63  &              \cr
$f_1 \eta        $       & 0.0033 &0.0032 &              \cr
$ \rho^+ \rho^-        $ & 4.30   & 16.8 &$ (<  9.5)    $ \cr
$ \rho^{\scriptscriptstyle 0} \rho^{\scriptscriptstyle 0}  $ & 1.04
& 2.20 &$0.4 \pm 0.3  $ \cr
$ \rho^{\scriptscriptstyle 0} \omega     $ & 2.13   & 3.94 &$3.9 \pm 0.6  $ \cr
$ \omega    \omega     $ & 1.07   & 1.62 &$1.4 \pm 0.6  $ \cr
$ \rho^\pm  a_1^\mp    $ & 0.099  &0.99  &                \cr
$ \rho^{\scriptscriptstyle 0} a_1^{\scriptscriptstyle 0}   $ & 0.028
& 0.14 &                \cr
$ \omega    a_1^{\scriptscriptstyle 0}   $ & 3.76   & 2.26 &                \cr
$ \rho^{\scriptscriptstyle 0} f_1        $ & 0.039  & 0.029&                \cr
$a_2^\pm \pi^\mp       $ & 0.88   & 0.74 & $1.3 - 2.6   $ \cr
$ a_2^{\scriptscriptstyle 0} \pi^{\scriptscriptstyle 0}    $ & 0.14
& 0.18 &                \cr
$ f_2\eta              $ & 0.0026 &0.0032&                \cr
$ a_2^{\scriptscriptstyle 0} \eta        $ & 0.0025 &0.0026&                \cr
$ f_2 \pi^{\scriptscriptstyle 0}         $ & 0.068  & 0.078& $0.41\pm0.12 $ \cr
$ a_2^\pm   \rho^\mp   $ & 0.040  & 0.028&                \cr
$ a_2^{\scriptscriptstyle 0} \rho^{\scriptscriptstyle 0}   $ & 0.013
&0.0052&                \cr
$ f_2 \rho^{\scriptscriptstyle 0}        $ & 0.067  &0.071 &$1.57\pm0.34  $ \cr
$ f_2 \omega           $ & 0.14   & 0.10 &$ 3.05\pm0.31 $ \cr
$ K^+ K^-              $ &  0.065 & 0.095&$0.1 \pm 0.01 $ \cr
$ K^{\scriptscriptstyle 0}\overline{K}^{\scriptscriptstyle 0}    $ &
0.0041& 0.024&$0.08\pm 0.01 $ \cr
$K^{\pm}K^{*\mp}       $ &  0.050 & 0.46 &$0.1 \pm 0.016$ \cr
$\,K^{\scriptscriptstyle 0} \overline{K}^{*{\scriptscriptstyle 0}}
                       $ &        &      &                \cr
/$ K^{*{\scriptscriptstyle 0}}K^{\scriptscriptstyle 0}
                       $ &  0.0044& 0.025&$0.12\pm0.02  $ \cr
$ K^{*+}   K^{*-}      $ &  0.055 & 0.19 &$0.13\pm0.05  $ \cr
$K^{*{\scriptscriptstyle 0}}\overline{K}^{*{\scriptscriptstyle 0}}$&
0.023  &0.035 &$0.26\pm0.05  $ \cr
\noalign{\hrule}
$ \Sigma               $ &  23.77 &45.96 &$30.94\pm3.91 $ \cr
\end{tabular}
\end{table}

\begin{table} [t]
\caption{
\label{TIV}
Ratios of branching ratios ``at rest'' for
some interesting channels as examples for dynamical selection rules.
The data are taken from Ref.\protect\cite{DGMF,AM}.
}
{

\begin{tabular}{c|ddd|c}
                                                               &D
&A(BOX)  &Born  &Exp. \cr
\noalign{\hrule}
$ {\overline{p}p\ ({^3S_1},I=0) \to \rho^\pm\pi^\mp} \over
  {\overline{p}p\ ({^1S_0},I=1) \to \rho^\pm\pi^\mp}$          &  4.67
& 2.29   & 3.45 &35$\pm$16\cr
$ {\overline{p}p\ ({^1P_1},I=0) \to \rho^\pm\pi^\mp} \over
  {\overline{p}p\ ({^3P_{1,2}},I=1) \to \rho^\pm\pi^\mp}$      &  0.34
& 0.11   & 0.19 &1.16$\pm$0.23\cr
\noalign{\hrule}
$ {\overline{p}p\ ({^1S_0},I=1) \to f_2\pi^{\scriptscriptstyle 0}}
\over
  {\overline{p}p\ ({^1S_0},I=1) \to \rho^\pm\pi^\mp}$          &  0.28
& 0.11   & 0.056 &$2.6        $\cr
$ {\overline{p}p\ ({^3P_1},I=1) \to f_2\pi^{\scriptscriptstyle 0}}
\over
  {\overline{p}p\ ({^3P_2},I=1) \to f_2\pi^{\scriptscriptstyle 0}}$
&  1.49  & 1.57   & 1.23 &$\approx 11 $\cr
\noalign{\hrule}
$ {\overline{p}p\ ({^1S_0},I=0) \to a_2^\pm\pi^\mp} \over
  {\overline{p}p\ ({^3S_1},I=1) \to a_2^\pm\pi^\mp}$           & 0.51
& 0.93   & 0.91 &$3.6 - 8    $\cr
\noalign{\hrule}
$ {\overline{p}p\ ({^1S_0},I=0) \to \rho^{\scriptscriptstyle
0}\rho^{\scriptscriptstyle 0}} \over
  {\overline{p}p\ ({^1S_0},I=0) \to \omega\omega }$            &  6.30
& 62.40  & 1.76 &$0.1 - 0.3   $\cr
$ {\overline{p}p\ ({^1S_0},I=1) \to \rho^{\scriptscriptstyle 0}\omega}
\over
  {\overline{p}p\ ({^1S_0},I=0) \to \omega\omega }$            &
13.17& 4.23   & 4.33 &$0.8         $\cr
$ {\overline{p}p\ ({^3S_1},I=1) \to \rho^{\scriptscriptstyle 0}\eta
}\over
  {\overline{p}p\ ({^3S_1},I=0) \to \omega\eta}$               &  0.47
& 10.95  & 0.98 &$ 0.55\pm0.12$\cr
\end{tabular}
}
\end{table}


\begin{thebibliography}{99}

\bibitem{AM}{C.\ Amsler and F.\ Myhrer,
{\sl Ann.\ Rev.\ Nucl.\ Part.\ Sci.} {\bf 41}\ (1991)\ 219,
(edited by
J.D.\ Jackson, H.E.\ Gove and R.F.\ Schwitters),
Annual Reviews, Inc., Palo Alto.}
%
\bibitem{DGMF}{C.B.\ Dover, T.\ Gutsche, M.\ Maruyama, and A.\ F\"a\ss ler,
{\sl Prog.\ Part.\ Nucl.\ Phys.\ }{\bf 29}\ (1992)87.}

\bibitem{Nijmegen}{P.H.\ Timmers, W.A.\ van der Sanden, and J.J.\ de Swart,
{{\sl Phys.\ Rev.\ }{\bf D 29} (1984) {1928}};
R.\ Timmermans, Th.A.\ Rijken, and J.J.\ de Swart,
{{\sl Phys.\ Rev.\ }{\bf C 50} (1994) {48}}.}

\bibitem{Paris}{M.\ Pignone, M.\ Lacombe, B.\ Loiseau, and R.\ Vinh Mau,
{{\sl Phys.\ Rev.\ }{\bf C 50} (1994) {2710}}.}

\bibitem{NNpirho}{G. Janssen, K. Holinde, and J. Speth,
{{\sl Phys.\ Rev.\ Lett.\ }{\bf 73} (1994) {1332}}.}
%
\bibitem{MHE}
{R.Machleidt, K.Holinde, Ch.Elster, {\sl Phys.\ Rep.\ } {\bf 149} (1987) 1.}
\bibitem{NNbI}
{T.\ Hippchen, J.\ Haidenbauer, K.\ Holinde,and V.\ Mull,\ \
{{\sl Phys.\ Rev.\ }{\bf C 44} (1991) {1323}}.}
%
\bibitem{Holz}
B.\ Holzenkamp, K.\ Holinde, and J.\ Speth,\ \
{{\sl Nucl.\ Phys.\ }{\bf A 500} (1989) {485}}.
%
\bibitem{NNbII}
{V.Mull, J.\ Haidenbauer, T.\ Hippchen, and K.\ Holinde,\ \
{{\sl Phys.\ Rev.\ }{\bf C 44} (1991) {1337}}.}
%
\bibitem{TabakinLiu}{G.Q. Liu and F. Tabakin,
{{\sl Phys.\ Rev.\ }{\bf C 41} (1990) {665}}.}

%
 \bibitem{Dumb}
 {O.\ Dumbrais et al., {{\sl Nucl.\ Phys.\ }{\bf B 216} (1983) {277}}.}

\bibitem{KunneII}{R.A.\ Kunne et al., {{\sl Phys.\ Lett.\ }{\bf B 261} (1991)
{191}}.}
%
\bibitem{Bradamante}
{A.\ Ahmidouch, PhD-Thesis, University of Gen\'eve, These No. 2672, 1994.}

\bibitem{Hasan}
{A.\ Hasan et al., {{\sl Nucl.\ Phys.\ }{\bf B 378} (1992) {3}}.}

\bibitem{pipiFSI} {V.\ Mull, K.\ Holinde, and J.\ Speth, {{\sl Phys.\ Lett.\
}{\bf B 275} (1992) {12}}.}
%
\bibitem{MullDiss}
{V.\ Mull, PhD-Thesis, Universit\"at Bonn 1993; \\
Berichte des Forschungszentrums J\"ulich Nr.\ 2844, 1993.}
\bibitem{BruecknerIV}{W.\ Br\"uckner et al.,\ \ {{\sl Phys.\ Lett.\ }{\bf B
166}
(1986) {113}}.}

\bibitem{KiI}{M.\ Kimura et al., {{\sl Nuov.\ Cim.\ }{\bf A 71} (1982) {438}}.}

\bibitem{KunneI}{R.\ A.\ Kunne et al., \ {{\sl Phys.\ Lett.\ }{\bf B 206}
(1988)
           {557}}; R.\ A.\ Kunne et al., \ {{\sl Nucl.\ Phys.\ }{\bf B 323}
           (1989) {1}}.}

\bibitem{Bertini}{R.\ Bertini et al., CERN-EP/89-83; F.\ Perrot-Kunne et al.,
{\sl`Proceedings of the 1st Biennial Conference on Low Energy Antiproton
Physics'}, Stockholm, 1990, eds.\ P.\ Carlson, A.\ Kerek and S.\ Szilagyi,
(World-Scientific, Singapore, 1991), p.251.\ }

\bibitem{KageyamaI} {T.\ Kageyama, T.\ Fujii, K.\ Nakamura, F.\ Sai, S.\
Saka\-moto, S.\ Sato, T.\ Takahashi, T.\ Tanimori, and S.\ S.\ Yamamoto, \
{{\sl
Phys.\ Rev.\ }{\bf D 35} (1987) {2655}}.}

\bibitem{NakII}
{K.\ Nakamura, T.\ Fujii, T.\ Kageyama, F.\ Sai, S.\ Saka\-moto, S.\ Sato,
T.\ Takahashi, T.\ Tanimori, S.\ S.\ Yamamoto, and Y.\ Takada,
\ {{\sl Phys.\ Rev.\ Lett.\ }{\bf 53} (1984) {885}}.}

\bibitem{BruecknerII}{W.\ Br\"uckner et al.,\ \ {{\sl Phys.\ Lett.\ }{\bf B
169}
(1986) {302}}.}

\bibitem{Birsa}{R.\ Birsa et al., \ {{\sl Phys.\ Lett.\ }{\bf B 246} (1990)
{267}}.}
\bibitem{BardinII}  {G.\ Bardin et al., \
in:{\sl` Proceedings of the 1st Biennial Conference
on Low Energy Antiproton Physics'},
Stockholm, 1990, eds.\ P.\ Carlson, A.\ Kerek and S.\ Szilagyi
(World Scientific, Singapore,1991) p.173.}
%
\bibitem{Sai} {F.\ Sai, S.\ Sakamoto , and S.S.\ Yamamoto,\ \ {{\sl Nucl.\
Phys.\ }{\bf B 213} (1983) {371}}.}
%

\end{thebibliography}
\end{document}